\documentclass[epj]{svjour}
\usepackage{latexsym}
\usepackage{graphics}
\usepackage{epstopdf}
\usepackage{amsmath}
\usepackage{mathtools}
\usepackage{braket}
\usepackage{graphicx}
\usepackage{dcolumn}
\usepackage{bm}
\usepackage{filecontents}

\begin{document}
\title{Quantum neural networks driven by information reservoir}
\titlerunning{Quantum neural networks}
\author{Deniz T\"{u}rkpen\c{c}e \inst{1}, Tahir \c{C}etin Ak{\i}nc{\i} \inst{1} \and Serhat \c{S}eker\inst{1}
}                     
%
%
\institute{Department of Electrical Engineering, \.{I}stanbul Technical University, 34469 \.{I}stanbul, Turkey}
\date{Received: date / Revised version: date}
%
\abstract{
This study concerns with the dynamics of a quantum neural network unit  in order to examine the suitability of simple neural computing tasks. More specifically, we examine the dynamics of an interacting spin model chosen as a candidate of a quantum perceptron for closed and open quantum systems. We adopt a collisional model enables examining both Markovian and non-Markovian dynamics of the proposed quantum system. We show that our quantum neural network (QNN) unit has a stable output quantum state in contact with an environment carrying information content. By the performed numerical simulations one can compare the dynamics in the presence and absence of quantum memory effects. We find that our QNN unit is suitable for implementing general neural computing tasks in contact with a Markovian information  environment and quantum memory effects cause complications on the stability of the output state. 
} 

\maketitle
\section{Introduction}\label{intro}
Artificial neural networks and machine learning have attracted considerable interest parallel to popular discussions on artificial intelligence. Machine learning is a field of computer science  which gives computers the ability to learn without being explicitly programmed and its inception would be traced back to the early days of artificial neural networks
\cite{russell_artificial_2009,mcculloch_logical_1943,Rosenblatt58}. Neural networks are interconnected information processing structures mostly based on binary McCulloch-Pitts neurons \cite{mcculloch_logical_1943} inspired by the biological context. Hebb's learning rule which was based on the modification of the weights of the connected units also depends on the biological and neurophysiological basis~\cite{hebb_organization_2002}. Classical learning rules govern the dynamics in a statistical information environment which is defined in terms of probability density functions~\cite{hecht-nielsen_neurocomputing_1990}. Therefore formulations and constraints of the learning laws depend on the relations between the global and the local information environments of each processing element. 

The idea of using quantum systems as neural networks ~\cite{lewenstein_quantum_1991,kak_quantum_1995} triggered various studies ~\cite{lewenstein_quantum_1994,lagaris_artificial_1997,zak_quantum_1998,narayanan_quantum_2000,ventura_quantum_2000,gupta_quantum_2001} including the attempts implementing the QNN proposals in the context of quantum computation~\cite{gupta_quantum_2001,panella_neural_2011,zhou_quantum_2012,schuld_simulating_2015,banchi_quantum_2016}. However, quantum computation operates by linear, unitary and reversible logic gates on qubit states initialized as pure states in closed systems~\cite{nielsen_quantum_2011}. Moreover, closed quantum systems have highly non-equilibrium nature in contrast to the QNN  stable output state requirement~\cite{schuld_quest_2014}. Dissipative quantum computation~\cite{verstraete_quantum_2009} has been uncovered as a reasonable platform for stable QNN states~\cite{schuld_quest_2014}. Open quantum systems reach steady states by a non-unitary evolution during the interaction with a quantum reservoir~\cite{breuer_theory_2007} and defining QNNs as open quantum systems~\cite{altaisky_towards_2016} is another direction for quantum neural computing. 

Quantum computation paradigm attached an intense interest by the inventions of quantum algorithms capable of factoring large integer numbers in  polynomial time or having speed ups by quantum search algorithms over their classical analogues~\cite{nielsen_quantum_2011}. Therefore, it's a fair inference to expect the advantages of the quantum processors over their classical counterparts in which their working principles have been directly mapped from the classical ones. However, there are just a few quantum algorithms better than the classical ones and most of the remained quantum algorithms are variants of them. In order to develop new quantum algorithms or quantum processors inspired from the classical ones, one should overcome the classical intuitions and investigate the quantum nature of systems intended to be developed. In our study, we will not attempt to apply classical learning rules or some quantum recipes directly taken from classical neural computing protocols but analyse the quantum dynamics of systems often referred to as quantum neural network models. 
        
More particularly, we investigate the dynamics of a small QNN unit in contact with a quantum information reservoir. The notion `Information reservoir' was initially introduced in a classical manner~\cite{deffner_information_2013} and also studied for quantum systems~\cite{mandal_maxwells_2013,strasberg_quantum_2017}. We simulate the proposed quantum system up to three input nodes with different states and examine the unitary and non-unitary dynamics. We first present the unitary dynamics of the QNN unit and compare with its open system case.  We show that in the unitary case the quantum state of the output node periodically corresponds to the unit fidelity of a target state which is a coherent superposition of the input node states. However, when the input nodes contact with the reservoirs carrying quantum information content in the Markov approximation, output node reaches a steady state which is a weighted statistical mixture of specific reservoir states. We also find that the steady state of the output node may contain some amount of coherence enough to reveal pure quantum effects depending on the state of the reservoirs.

Our model diverges from the dissipative quantum computation model since we adopt a collisional model ~\cite{scarani_thermalizing_2002,ziman_diluting_2002} which enables simulating the non-unitary dynamics by both Markovian and non-Markovian ~\cite{cakmak_non-markovianity_2017} reservoir models. We observe back flow of quantum information from the reservoir to the the system and report a quantum memory effect of a QNN due to the non-Markovian dynamics.    
\section{Model and system dynamics}\label{sec:Sc1}
The goal of implementing advanced cognitive tasks revealed neural networks~\cite{hecht-nielsen_neurocomputing_1990}. Mathematically speaking, the simplest neural network unit is a perceptron which is a data classifier with $x_1, x_2\ldots x_N$ input nodes connected to an output node with corresponding adjustable weights $w_1,w_2,\ldots w_N$. The output node experience the weighted linear summation of the input information as $y=\sum_i x_iw_i$ modulated by an activation function $f(y)$ which returns an output depending on the value of $y$.

We adopt the quantum model of neural networks by replacing the nodes by identical two-level quantum systems (qubits) ~\cite{altaisky_towards_2016,diamantini_quantum_2006,kouda_qubit_2005}. The interactions between qubits modelled by dipole-dipole coupling. The system Hamiltonian reads 

\begin{equation}
H=\frac{\omega}{2}\sum_i^N \sigma_i^z+\left(\sum_i^{N-1} J_i \sigma_i^{+}\sigma_{out}^{-}+H.c.\right) 
\end{equation}
which is plausible for quantum effects also in biological systems~\cite{oreilly_non-classicality_2014}. Here, $\sigma^{\mp}$ are the Pauli raising and lowering operators for the qubits, $\sigma_{out}$ is the respective Pauli operator for output qubit and $j_i$ is the coupling coefficient between input quantum neurons and the output neuron. We employ density matrix formalism in our study to represent the quantum states of the system in question. The quantum neurons are initially  assumed to be in a product state as $\rho(0)=\rho_1(0)\otimes\rho_2(0)\otimes\ldots\otimes\rho_{out}(0)$. Individual qubit states were chosen  as $\rho(0)=\ket{+}\bra{+}$ in order to initially provide a blank memory. Here, $\ket{+}=\frac{1}{\sqrt{2}}\left(\ket{\uparrow}+\ket{\downarrow}\right)$ in which the states $\ket{\uparrow}, \ket{\downarrow}$ are orthogonal spin states known as computational basis in quantum computing language.   Since we also track the evolution of coherence to investigate the system dynamics in detail, we use $l_1$ norm of coherence~\cite{baumgratz_quantifying_2014}

\begin{equation}\label{eq:norm}
C_{l_1}(\rho)=\sum_{i\neq j} |\rho_{i,j}|
\end{equation}  
in order to quantify coherence. 
The closed quantum system dynamics are governed by Liouville-Von Neumann equation $\dot{\rho}=-i[H,\rho]$ or equivalently by a unitary evolution such as $\rho(t)=U_t\rho(0)U_t^{\dagger}$ where $U_t=e^{-iHt}$ is time shift operator and $H$ is a time independent Hamiltonian. Planck constant divided by $2\pi$ was taken $\hbar=1$ throughout the manuscript. Mostly we are interested in the dynamics of typical quantum observables $\langle \sigma_{\nu}(t)\rangle_{out}=\text{Tr}[\rho_{out}(t)\sigma_{\nu}]$ of the output neuron where $\nu:x,y,z$. One obtains the  quantum state of output neuron as $\rho_{out}(t)=\text{Tr}_i[\rho(t)]$ where $\text{Tr}_i$ stands for a partial trace operation over input neurons degrees of freedom. 

In our case dynamically we need to quantify how close the quantum state of the output neuron to a desired quantum state. To this end, we use quantum fidelity between two quantum states in the form

\begin{equation}\label{eq:Fid}
\mathcal{F}(\rho,\sigma)=\text{Tr}\sqrt{\sqrt{\rho}\sigma\sqrt{\rho}}.
\end{equation}

A quantum system in contact with a reservoir with large degrees of freedom can be defined as an open quantum system. As we stressed before, the objective of this paper is to investigate a quantum version of perceptron model by focusing on a QNN unit as an open quantum system. Open quantum system dynamics can be explained by a master equation 

\begin{equation}
\dot{\rho}=-i[H,\rho]+\mathcal{L}\left(\rho\right)
\end{equation}
with a non-unitary and irreversible evolution \cite{verstraete_quantum_2009,breuer_theory_2007}. Here,
\begin{equation}
\mathcal{L}\left(\rho\right)=\sum_k L_k\rho L_k^{\dagger}-\frac{1}{2}\lbrace L_k^{\dagger}L_k,\rbrace_{+}
\end{equation}
is a Liouvillian in Lindblad form and $L_k$ are the super-operators affecting locally to the subspaces of the system in question and $\lbrace . \rbrace_{+}$ stands for anti-commutation. The initial information of the open quantum system subjected to this dynamics is erased and the interaction with the reservoir drives the system to a steady state.  The evolution of the system to such a stationary state can also be referred to as equilibration in a long-term period.  In the above formulation, the state of the equilibrated system is independent of its initial state. In other words, the system does not remember its past hence the evolution is Markovian. However in this study, we adopt a collisional model to describe open system dynamics enabling  to explain by both Markov and non-Markov approaches beside from the statement describing the generic Lindblad model above. As explained in the Results section, collision model ~\cite{scarani_thermalizing_2002,ziman_diluting_2002,cakmak_non-markovianity_2017} defines open system dynamics by the interaction of the system in question with sub-environments (units) of a reservoir for finite time intervals. The type of interaction (inter-unit or only unit-system interactions) determines Markov or non-Markov character of the system modelling  and the open system dynamics can be obtained by tracing out the environmental degrees of freedom.

In this study, we limit the quantum states of the information reservoir~\cite{deffner_information_2013,mandal_maxwells_2013,strasberg_quantum_2017} to pure states of two-level systems. Geometrical Bloch sphere representation 
$\ket{\psi\left(\theta,\phi\right)}=\cos\left(\frac{\theta}{2}\right)\ket{0}+e^{i\phi}\sin\left(\frac{\theta}{2}\right)\ket{1}$ is well-known and useful for describing these states.  The underlying physical phenomenon of the study is the justification of einselection~\cite{zurek_decoherence_2003} which is a transformation process from pure states into mixtures of the preferred basis. This typical decoherence process of  open systems can also be discussed in terms of additive properties of different reservoirs~\cite{chan_quantum_2014} interacting with the same quantum system. We often use $\rho_m=\sum_i p_i\ket{\psi_i}\bra{\psi_i}$ as target states where $\ket{\psi_i}$ is the quantum state of the $i^{th}$ reservoir and the $\rho_m$ is the density matrix representing the statistical mixture of the reservoir states such that $\sum_i p_i=1$. We also investigate the unitary dynamics of the proposed QNN unit, therefore we also check the state of the output node by using the coherent superposition  states of the input nodes $\rho_c=\ket{\psi_C}\bra{\psi_C}$ as the target state where $\ket{\psi_C}=\sum_i c_i\ket{\psi_i}$ and $\sum_i |c_i|^2=1$. Temporal evolution of several observables were discussed for the proposed model analysis.  The notation we use for the fidelity expressed as follows. Mostly, we are interested in the quantum state of the output neuron with respect to $\rho_c$ or $\rho_m$ as target states therefore, formally we use $\mathcal{F}(\rho_c,\rho_{out})$ and $\mathcal{F}(\rho_m,\rho_{out})$ in our calculations. However, $\rho_{out}$ is not necessarily the only subsystem or $\rho_c$ and $\rho_m$  are not the only target states to be investigated. For instance we use,  $F01$, $F12$, $F+3$ or $F03$ to represent the fidelities $\mathcal{F}(\ket{\uparrow}\bra{\uparrow},\, \rho_1)$, $\mathcal{F}(\ket{\downarrow}\bra{\downarrow},\, \rho_2)$ , $\mathcal{F}(\ket{+}\bra{+},\, \rho_{out})$ and $\mathcal{F}(\ket{\uparrow}\bra{\uparrow},\, \rho_{out})$ respectively where $\rho_i$ is the $i^{th}$ node of the QNN unit. 

\section{Results}

In this section we present the numerical results of our QNN unit, depicted in the bottom of the left panel of Fig.~\ref{fig:Fig1}. First, we analyse the dynamics as a closed quantum system by a unitary evolution and then connect the input quantum neurons with quantum reservoirs carrying one qubit information. 
\subsection{Unitary dynamics}

\begin{figure*}[!t]
\includegraphics[width=6.3in]{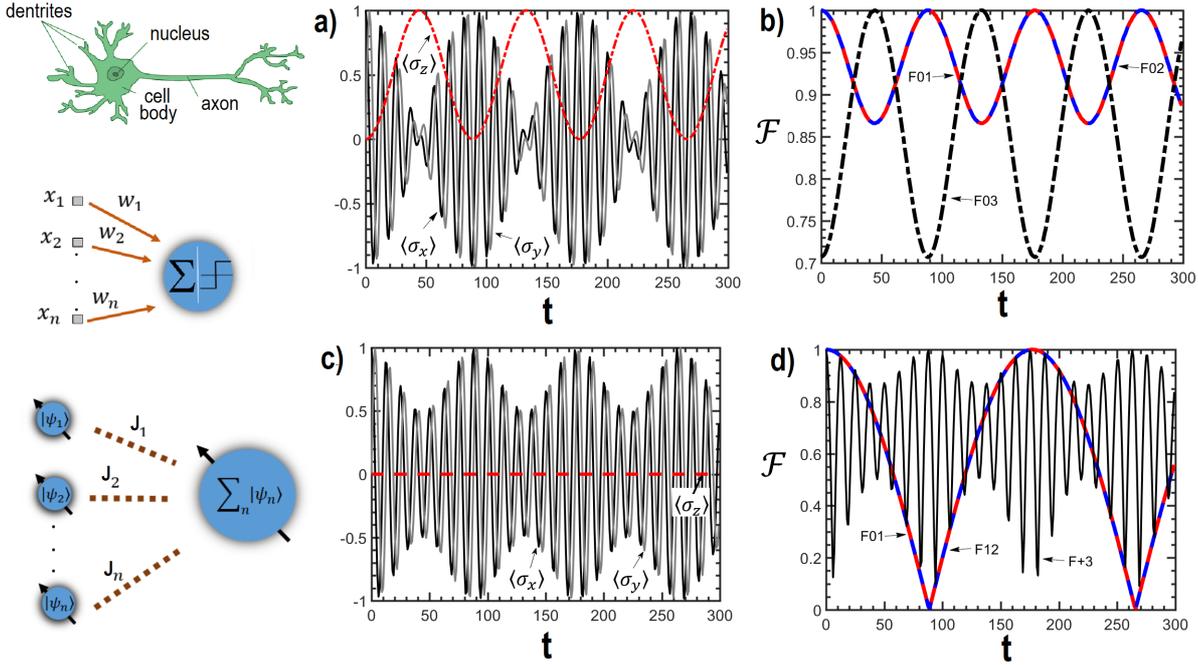}
\caption{ \label{fig:Fig1} A biological neuron (top of right panel), a perceptron (middle of the left panel) , a QNN unit with $n$ input nodes (bottom of the left panel) and unitary dynamics of a QNN unit with two input nodes. Output node was prepared in the $\ket{+}$ state. The coupling coefficient between two input nodes and the output node is $J_1=J_2=0.05$. In (a), time evolution of observables $\langle \sigma_x \rangle$, $\langle \sigma_y \rangle$  and $\langle \sigma_z \rangle$ of the output node was depicted. Initial states of both input nodes were prepared as $\ket{\uparrow}$ in (a). In (b), evolution of the fidelity between the quantum states of input modes with respect to their initial states $F01$, $F02$ and the quantum state of the output mode with respect to $\ket{\uparrow}$. In (c), the same observables were depicted for the output node. Here, the initial states $\ket{\uparrow}$ and $\ket{\downarrow}$ respectively were chosen for the first and the second input node and the initial state of the output node is $\ket{+}$. The coupling coefficient $J$ and time is dimensionless and scaled by the Bohr frequency $\omega$.} 
\end{figure*}

A weakly interacting spin model can be examined to investigate the extent to which a quantum version of a perceptron can be represented by. In this model, the goal is to find the optimal unitary operator $U$ in order to reach the desired output state with a minimum cost function~\cite{lewenstein_quantum_1994}. A practical figure of merit is the spin polarisation $\langle \sigma_z \rangle$  for the output neuron quantum state ranges between 1 and -1. To this end, we calculate the reduced dynamics of the system. In Fig.~\ref{fig:Fig1} we consider the model with two input nodes with specific quantum states. We observe typical oscillatory temporal dynamics of this closed system. In Fig.~\ref{fig:Fig1} (a) we set the both input nodes in $\ket{\uparrow}$ states and in Fig.~\ref{fig:Fig1} (c) we set the initial quantum state of the input nodes as two orthogonal quantum states $\ket{\uparrow}$ and $\ket{\downarrow}$ respectively. In the former case spin polarisation of the output node oscillates between 0 and 1. In the latter, spin polarisation returns 0 throughout the evolution corresponding to the two orthogonal input state case with the same $J$ coupling values. 

As clear in Figs.~\ref{fig:Fig1} (b), (d) both input and output states are not stable  due to the nature of unitary dynamics.
The quantum states of the output node and the input nodes as well as their observables present the oscillatory dynamics. In Fig.~\ref{fig:Fig1} (b) both the initial quantum states of the input nodes are $\ket{\uparrow}$ and it's assumed that this state is also the desired output state. We note that an appropriate unitary operator $U$ can be constructed by this dynamics with an available operation time $t$ corresponding to the fidelity of  the output state $F03=1$. 
In Fig.~\ref{fig:Fig1} (d) the two input nodes were chosen as orthogonal  $\ket{\uparrow}$ and  $\ket{\downarrow}$ states and in this case the linear combination of these states $\ket{+}$ was assumed to be the desired output quantum state. In this case, though the initial state of the output node is $\ket{+}$ which is already the desired output state,             
obtaining $U$ corresponding to $F+3=1$ looks harder due to the more complex oscillatory dynamics. Given input states, the typical parameters will be the $J_i$ couplings and the evolution time $t$ to reach the optimal $U$ for a desired output state. Inclusion of the correct interaction time as well as $J_i$ values  makes the situation complicated for closed systems particularly for large number of input nodes. Moreover, closed systems are not realistic (for both quantum and classical cases) and the most straightforward way to obtain a stable quantum state is to contact the system with a reservoir. Then dealing with QNNs inevitably becomes an open quantum system problem.

\subsection{Markovian dynamics}\label{sec:Sc2}
Understanding the field of open quantum systems is of importance for quantum control~\cite{leibfried_quantum_2003,raimond_manipulating_2001}, preserving system from undesired environmental noise~\cite{suter_colloquium_2016} or engineering system environment interactions~\cite{verstraete_quantum_2009}.  Particularly for small quantum systems it's been shown that any quantum observable will equilibrate to a steady value, given soft assumptions about the initial conditions of the system~\cite{reimann_foundation_2008,linden_quantum_2009}. 

In this subsection, we analyse the QNN unit as a system that interacts weakly with an information reservoir in the Markov approximation. In these interactions, the quantum state of the system irreversibly lost through the reservoir degrees of freedom and the temporal evolution of the system only depends on the current state of the system. On the other hand, recent studies show that  quantum environments are not just the waste bins in which the connected system information is lost, but they can also be defined as communication channels in which they transmit information they retain to the connected systems~\cite{blume-kohout_simple_2005,zwolak_redundancy_2017}. For instance, definition of conventional quantum thermalization states that under some conditions a thermal reservoir in which a quantum system interacts with, sends the temperature information to the system at its steady state~\cite{liao_single-particle_2010,turkpence_quantum_2017}. 

In our case, a quantum information reservoir weakly interacts with the local subsystems (input nodes) of the QNN unit. We show that time evolution of the expectation values of typical quantum observables dynamically converge to steady values in accordance with the quantum state of the reservoir. By the rapid interest to quantum thermodynamics~\cite{quan_quantum_2007,kieu_second_2004,hardal_superradiant_2015,altintas_quantum_2014,altintas_rabi_2015,turkpence_quantum_2016,turkpence_photonic_2017} in the past decade, the need for classification  of quantum reservoirs has emerged. In this context, quantum reservoirs classified as work, heat and information reservoirs~\cite{strasberg_quantum_2017}. 

\begin{figure*}[!t]
\includegraphics[width=6.2in]{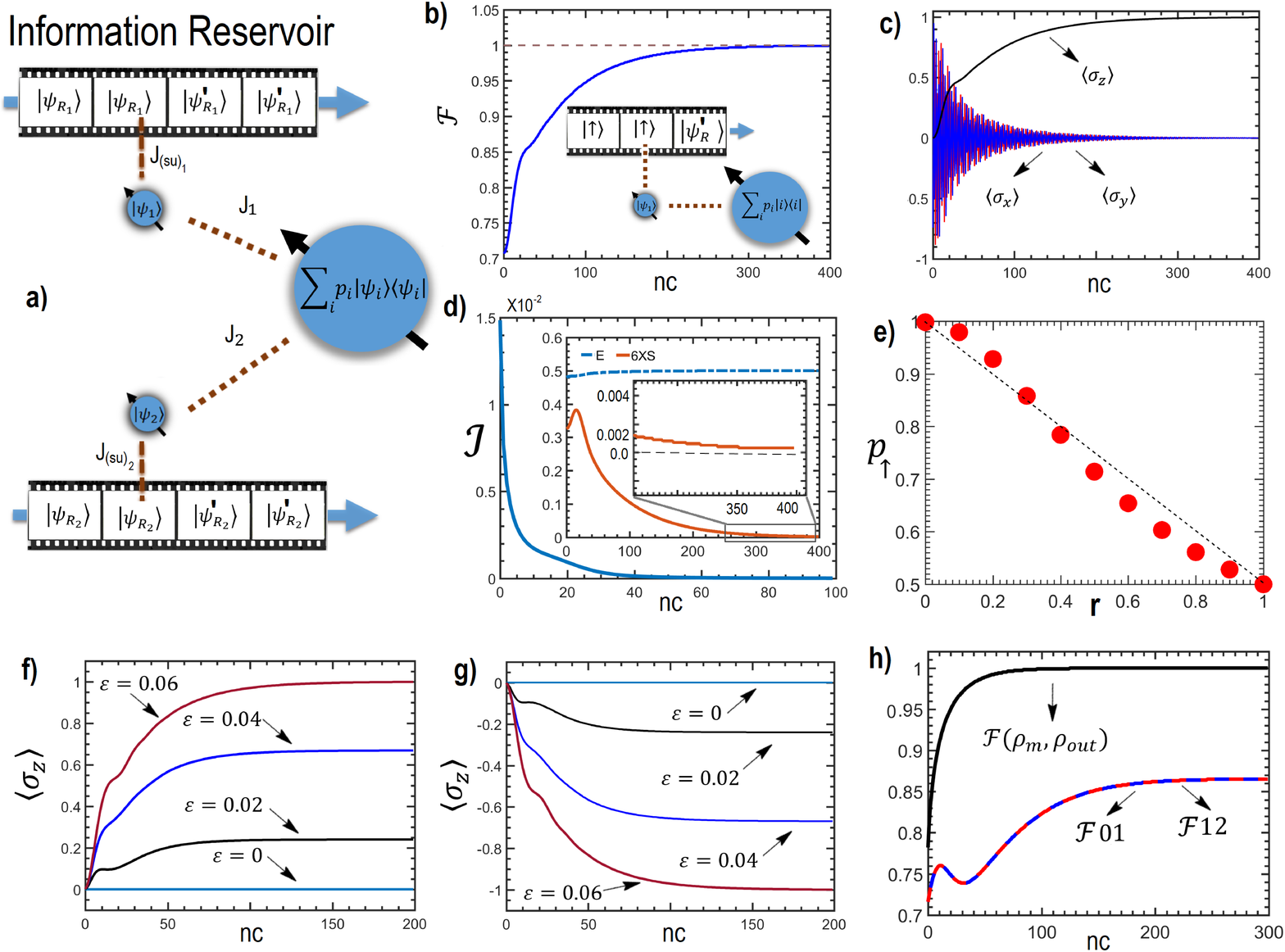}
\caption{\label{fig:Fig2} QNN unit dynamics in the presence of information reservoirs. (b) Evolution of fidelity between the quantum state of one of the incoming units and the state of the output node $\rho_{out}$ of the perceptron depending on the number of collisions (nc). Coupling between input node and the units (between the system and the units) $J_{su}$ is equal to the coupling between the input-output node $J=0.05$. (c) Evolution of observables of the $\rho_{out}$ during the contact with the reservoir. (d) Evolution of quantum mutual information between an interacted unit and $\rho_{out}$. Also the evolution of Von-Neumann entropy $S$ and the internal energy $E$ plotted (inset). (e-h) Quantum dynamics of several quantities of the QNN unit with two input nodes connected to information reservoirs with two orthogonal quantum states. (e) Dependence of the higher energy level population $p_{\uparrow}$ of output node for unequal input node-resevoir coupling  $J_{(su)_1} \neq J_{(su)_2}$  to a parameter $r$. Here, input nodes-output node couplings are equal and $J=0.06$. On the other hand, $J_{(su)_1}=J$ and  $J_{(su)_2}=Jr$. The evolution of spin polarizations (f), (g) and fidelities (h) for equal inter nodes coupling $J=0.06$ and system reservoir coupling  $J_{(su)_1}=J_{(su)_2}=J$. In (h), $\mathcal{F}(\rho_m, \rho_{out})$ is the fidelity between the state of the output node with respect to the mixture of two orthogonal reservoir states, $\mathcal{F}01$ and $\mathcal{F}12$ are the fidelities between the first input node and the second input node with respect to the quantum state of the reservoir they contact. Here, the coupling of the first node to the out put node is equal to the coupling of the second node to the out put node  $J_1=J_2=0.06$ In (g), $J_1=J$ and $J_2=J-\varepsilon$ while in (h) $J_1=J-\varepsilon$ and $J_2=J$ where $J=0.06$. Coupling of the nodes to the reservoir units is the same and $J=0.06$ for (f), (g) and (h). The interaction time between the nodes and units is $\tau=0.1\pi/J$ for all subplots.} 
\end{figure*}

As we stressed above, we model open system dynamics by a collisional model in which the reservoir was modelled by repeated interactions between the QNN system and statistically independent and identically prepared units~\cite{strasberg_quantum_2017,cakmak_non-markovianity_2017}.
Fig.~\ref{fig:Fig2} (a) illustrates this scheme by a strip of qubit strings (units) interacting sequentially with the input nodes of the QNN unit. The interaction Hamiltonian between environment units and a single input node is in the same form between the input nodes and the output node as
 \begin{equation}
 H_{su}=J_{su}\sigma_s^{+}\sigma_u^{-}+H.c..
\end{equation}  
In the scenario, input nodes interact with each sub-environment unit for a time $\tau$ then disconnect and interact with the forthcoming unit. The propagator for each interaction is $U_{su}=\text{exp}(-iH_{su}\tau)$. For the Markovian  reservoir case the units have no interaction each other. After each interaction, the system of interest obtained by tracing out the environmental degrees of freedom

\begin{equation}\label{eq:Tru}
\rho_s(t+\tau)=\text{Tr}_u\left[U_{su}(\tau)\rho_{su}(t)U_{su}^{\dagger}(\tau)\right].
\end{equation}

It's shown that this kind of discrete time evolution is equivalent to a master equation approach in terms of divisibility of quantum channels~\cite{wolf_dividing_2008}. Following these specifications, we first connect the reservoir to the QNN unit with a single node which is the simplest version. Throughout the calculations, initial states of the input and the output nodes are $\ket{+}$  unless the opposite stated. Fig.~\ref{fig:Fig2} (b) depicts the evolution of the fidelity between output node and the reservoir unit quantum state 
$\mathcal{F}(\ket{\uparrow}\bra{\uparrow},\rho_{out})$ depending on the number of collisions between units and input node. The state of the output node reaches the unit fidelity with the reservoir quantum state and the spin polarization points up at the end of the evolution (Fig.~\ref{fig:Fig2} (c)) while the evolution of $\langle \sigma_x \rangle$ and $\langle \sigma_y \rangle$ represent the decay of the off-diagonals. This constitutes a typical example of equilibration of an open quantum system. 

Information is inevitably related with thermodynamics~\cite{landauer_irreversibility_1961,mandal_work_2012}. Therefore we also analyse some thermodynamic quantities in this model. For instance, Von Neumann entropy
\begin{equation}
S(\rho)=-\text{Tr}[\rho ln\rho]
\end{equation}
quantifies the amount of uncertainty in a quantum system represented by a density matrix. On the other hand, internal energy of a quantum system can be represented by 
\begin{equation}
E=\text{Tr}[ \rho H]
\end{equation}
where $H$ is the Hamiltonian of the system. Inset of Fig.~\ref{fig:Fig2} (d) depicts the evolution of the internal energy $E$ and the Von Neumann entropy $S$ of the reservoir units with an appropriate scaling after the interaction with the QNN unit input nodes. This evolution can be modelled by using Eq.~(\ref{eq:Tru}) but by tracing out the QNN unit this time. In this evolution one can see that $E$ rapidly saturates while $S$ reduces to a very small non-zero value thus satisfying
\begin{equation}
\frac{\Delta E_u}{\Delta S_u}\rightarrow 0
\end{equation}
 which confirms the reservoir under consideration is an information reservoir~\cite{strasberg_quantum_2017}. 
Another useful measure is quantum mutual information 
\begin{equation}
\mathcal{I}(\rho_s:\rho_u)=S(\rho_{s})+S(\rho_{u})-S(\rho_{su})
\end{equation}
defined in terms of Von Neumann entropy, quantifies the classical and quantum correlations between subsystems~\cite{nielsen_quantum_2011}.
Novel approaches interpret $\mathcal{I}$ as a measure of `learned' information by considering the amount of uncertainty  reduced of $\rho_s$ by knowing $\rho_{u}$~\cite{goldt_stochastic_2017}. Fig.~\ref{fig:Fig2} (d) shows the decrease of quantum mutual information between output node and the information reservoir unit after each interaction.   
This oversimplified version of a QNN unit with a single input node can be attributed to a supervised learning scheme with an already trained system by a single labelled data $\ket{\uparrow}$. In this task, spin polarization reaches its maximum value and the QNN unit should fire an `action potential' since the desired information reaches the output node. However in this study, we did not attempt to formulate a quantum activation function which can be a topic of an else study. 

\begin{figure*}[!t]
\includegraphics[width=6.2in]{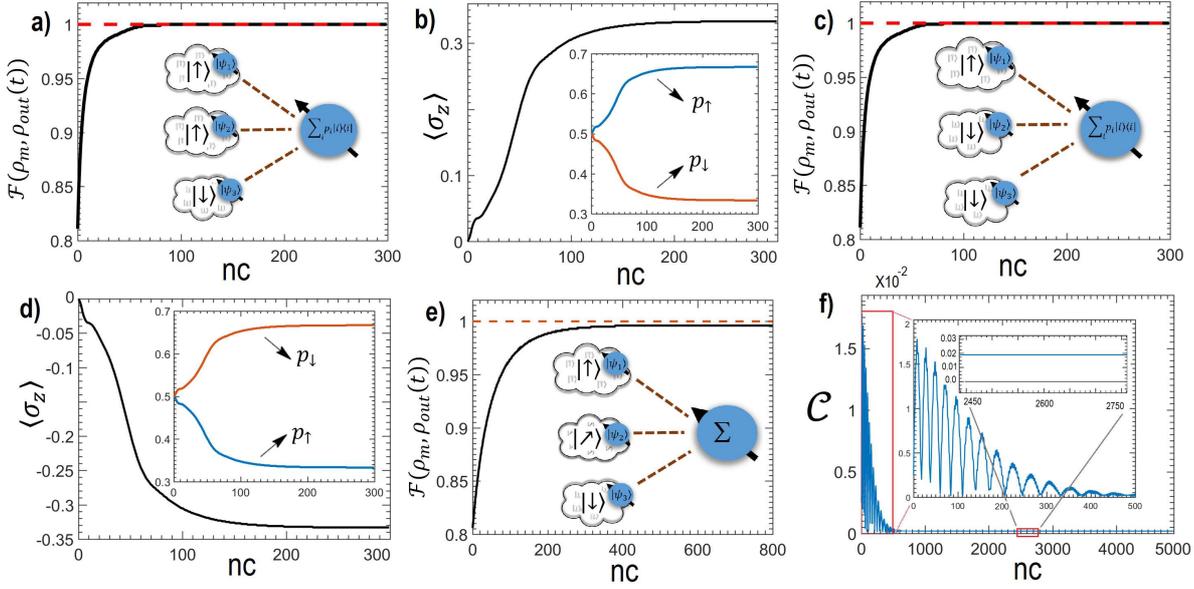}
\caption{\label{fig:Fig3}  Dynamics of the QNN unit with three input nodes in contact with various information reservoirs. (a) Evolution of the output node fidelity  with respect to the mixture of two $\ket{\uparrow}$ and one $\ket{\downarrow}$ states and (b) evolution of the spin polarization and energy level populations of the output node (while the interaction of the three input nodes with three information reservoirs of two $\ket{\uparrow}$ and one $\ket{\downarrow}$ states) depending on the number of collisions between reservoir units and the input nodes. (c) Evolution of the output node fidelity with respect to the mixture of two $\ket{\downarrow}$ and one $\ket{\uparrow}$ states and (d) evolution of the spin polarization and energy level populations of the output node (while the interaction of the three input nodes with three information reservoirs of two $\ket{\downarrow}$ and one $\ket{\uparrow}$ states) depending on the number of collisions between reservoir units and the input nodes. (e) Evolution of the output node fidelity with respect to the mixture of $\ket{\uparrow}$, $\ket{\nearrow}$ and $\ket{\downarrow}$ states and the evolution of the $l_1$ norm of coherence of the output node (while the interaction of the three input nodes with three information reservoirs, respectively $\ket{\uparrow}$, $\ket{\nearrow}$ and $\ket{\downarrow}$)  depending on the number of collisions between reservoir units and the input nodes. Initial states of the output nodes are $\ket{+}$ in (a)-(d) and $\ket{\uparrow}$ in (e)-(f). Reservoir units and input nodes $J_{(su)}$ couplings and inter node couplings are equal and $J=0.05$. The interaction time between the nodes and units is $\tau=0.1\pi/J$ for all subplots.} 
\end{figure*}

Next, we add a second input node as in the closed system case and connect each of them to distinct information reservoirs in Figs.~\ref{fig:Fig2} (e)-(h). In this case, the stability of the output quantum state is also apparent as in the single node case equilibration process. This time however, as in Fig.~\ref{fig:Fig2} (h) the output state reaches a stable value with unit fidelity with a target stage as a statistical mixture of quantum information coming from the quantum reservoirs coupled to the input nodes. We also equilibrate the system  with two reservoirs again consisting orthogonal quantum states but with unequal weights with $J_{(su)_1}\neq J_{(su)_2}$ as in Fig.~\ref{fig:Fig2} (e) and with $J_1\neq J_2$ as in Figs.~\ref{fig:Fig2} (f), (g). In Fig.~\ref{fig:Fig2} (e) higher level population of the output node  $p_{\uparrow}$ reveals an almost linear dependence with the control parameter $r$ reflecting the inequality of the couplings. This is of importance to the dynamical control of the QNNs in the context of implementing neural computing tasks. One can observe a similar control process on the polarization of the output node as in Figs.~\ref{fig:Fig2} (f), (g) with $\varepsilon$.  The polarization state of the output node ranges between 1 and -1 and reach these values in two limit cases for $\varepsilon=J$. This shows that the implementing the quantum version of a training process by adjusting the weights (couplings) by a teacher according to the feedback of the information coming from the output neuron is possible in a stable manner in the presence of the quantum information reservoirs.  

We extend our discussion by adding a third input node to our QNN unit interacting with a third information reservoir as in Fig.~\ref{fig:Fig3}. In this case we set all the couplings equal and we do not modify them since this time we would like to focus on the effect of the number of the input nodes and the quantum state of the information reservoirs coupled to these nodes. Since the output node is a two-level system, the steady quantum state of the output node is always in the form $\rho_m=p_{\uparrow}\ket{\uparrow}\bra{\uparrow}+p_{\downarrow}\ket{\downarrow}\bra{\downarrow}$ such that $p_{\uparrow}+p_{\downarrow}=1$ no matter how large the number of the reservoirs are. The populations $p_{\uparrow}$ and $p_{\downarrow}$ explicitly depend on the number of the orthogonal reservoirs. Numerically we observe that difference in the energy level populations, hence the spin polarization of the output node at steady state depends on the unequal number of information reservoirs for the orthogonal quantum states coupled to the input nodes. More generally, with our results one can define the steady state of the output node in terms of the number of orthogonal state reservoirs as
\begin{equation}\label{eq:EQpoint}
\rho_{out}=\frac{N_{\uparrow}}{N}\Pi_{\uparrow}+\frac{N_{\downarrow}}{N}\Pi_{\downarrow}
\end{equation}
where $\Pi_i=\ket{i}\bra{i}$ is the projector of the `selected' basis (pointer states) such that $i\in\{\uparrow,\downarrow\}$. Here, $N_{\uparrow}$ is the number of input nodes coupled to $\ket{\uparrow}$ state reservoirs while $N_{\downarrow}$ is the number of input nodes coupled to the $\ket{\downarrow}$ state reservoirs where $N=N_{\uparrow}+N_{\downarrow}$ is the total number of reservoirs input nodes interact. For instance as in Fig.~\ref{fig:Fig3} (b), higher level population  is $p_{\uparrow}\sim$~0.667 when the input nodes couple with two $\ket{\uparrow}$ and one $\ket{\downarrow}$ reservoirs. On the other hand higher level population is (see Fig.~\ref{fig:Fig3} (d)) $p_{\uparrow}\sim$~0.333 when the situation is opposite. This numerically confirm the validity of Eq.~\ref{eq:EQpoint} in Figs.~\ref{fig:Fig2} (a), (c) by obtaining the unit fidelity of the output node state for the $\rho_m$ statistical mixture of the reservoir states as the target state. Note that Eq.~\ref{eq:EQpoint} holds only for the reservoir $\ket{\uparrow}$ and $\ket{\downarrow}$ states. A discussion about the inclusion of coherent reservoir states will be given below. 

By now , we presented the reservoir induced stability of the output states by the pointer states which could be evaluated in the context of einselection phenomenon. \cite{zurek_decoherence_2003,blume-kohout_simple_2005,zwolak_redundancy_2017}.
Einselection is referred to as a decoherence-induced selection process of the set of pointer states remains stable during the system-environment interaction process. In Figs.~\ref{fig:Fig2} and ~\ref{fig:Fig3} we show that stable output states can be pure or mixed depending on the control parameters. In other words, we show that robust quantum neural computing tasks can be possible by an engineered environmental monitoring for a QNN unit. In fact, evolution from pure states towards mixed states is not surprising for the quantum systems in contact with the reservoirs. However, these results become important for our QNN unit since we characterize the situation in terms of control parameters that would be useful for quantum versions of neuro-computing tasks. Though mixed states may seem useless for exploiting quantum resources, it's been shown that mixed states can successfully simulate pure states and implement quantum quantum computing tasks called mixed state quantum computing~\cite{cory_ensemble_1997,knill_power_1998,vidal_efficient_2003}.

\begin{figure*}[!b]
\includegraphics[width=6.2in]{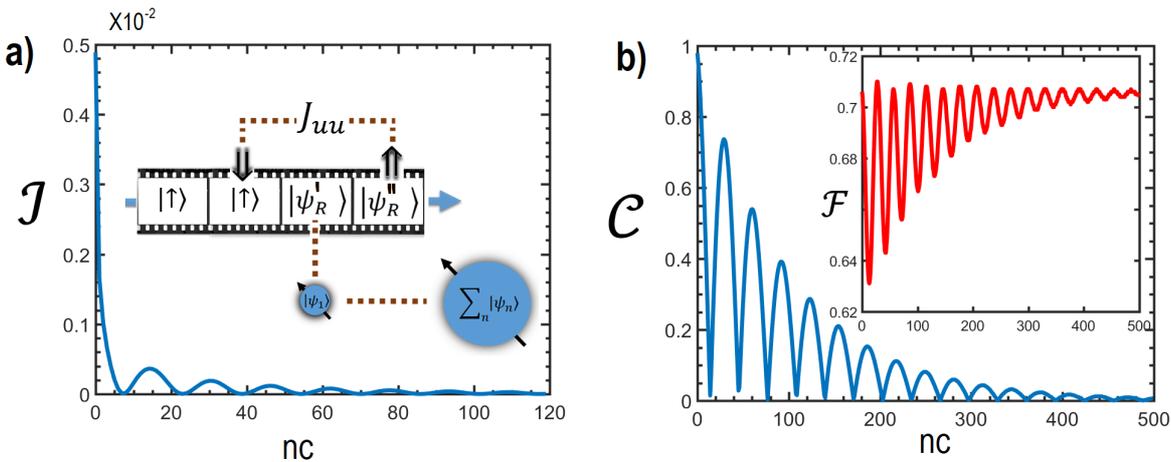}
\caption{\label{fig:Fig4} Dynamics of the one input node QNN unit coupled to a Non-Markov reservoir. (a) Mutual information dynamics between the output node of the QNN unit and the reservoir. (b) Evolution of $l_1$ norm of coherence of the output node and the fidelity (inset) between the state of the output node and the state of the incoming  units. During these evolutions input node-reservoir unit coupling and input node-output node coupling is equal and $J=0.5$. On the other hand inter-units coupling $J_{uu}=0.25$. Node-unit interaction time is $\tau_{su}=0.05/J$ and inter-units interaction time is $\tau_{su}=(\pi/4)/J_{uu}$.} 
\end{figure*}
Since pure states were assumed for information reservoirs in our study, any point on the Bloch sphere should be allowed as possible states of the information reservoirs  in contact with the input nodes. To this end, we also take an arbitrary pure state as the quantum state of the information reservoir. We choose $\ket{\psi(\theta)}\equiv \ket{\nearrow}$ for $\theta=\pi/6$ as the quantum state of the third reservoir for our numerical calculations. Note that this state has some amount of coherence thus we also present the dynamics of the QNN unit in contact with a coherent reservoir. In Figs.~\ref{fig:Fig3} (e) and (f) the initial states are $\ket{\uparrow}$ in order to begin the process with zero quantum coherence. We depict the relevant results in Figs.~\ref{fig:Fig3} (e), (f). First, we would like to know if the state of the output node also in this case converges to an equally weighted mixture of the states of the reservoirs presented in Fig.~\ref{fig:Fig3} (e). According to our numerical results the fidelity is $0.996$ means that we can not reach unit fidelity as presented in Fig.~\ref{fig:Fig3} (e) where the target state is $\rho_m$ the mixture of the states $\{\ket{\uparrow}, \ket{\nearrow},\ket{\downarrow}\}$. 

Since we are very close to the unit fidelity quantitatively  we can still hold our statements qualitatively about the steady outputs  consisting mixtures of the states of the reservoirs including the coherent information reservoirs. 
But in our view, the most significant task is the coherence of the output node. As presented in Fig.~\ref{fig:Fig3} (f) there is a non-vanishing coherence of the output node in the long term limit  when the third node is in contact with the reservoir with the $\ket{\nearrow}$ state. This is known as `steady coherence'~\cite{li_steady_2015} and it's very precious for quantum information tasks. The coherence we calculate is $\mathcal{C}_{l_1}=1.8\times10^{-4}$. It's been shown that this amount of coherence is sufficient to reveal non-traditional quantum effects for three-level~\cite{scully_extracting_2003} or multi-level~\cite{turkpence_quantum_2016} quantum systems. The appearance of this steady coherence can also be considered in terms of  additive properties of independent reservoirs in Markov regime~\cite{chan_quantum_2014}. Since we assume that we can control the coupling to the reservoirs, it appears that our proposed QNN unit is capable of switching from classical information to quantum information by simply coupling to coherent reservoirs resulting steady coherence on the output node alongside the information reservoirs selecting the classical pointer states.  

\subsection{Non-Markovian dynamics}
In many cases the Markov approximations between the reservoir and the system is not valid. For the dynamical evolution of a system connected to non-Markovian reservoirs, one observes an amount of back-flow of information. This back action is mostly referred to as memory effects of the system~\cite{de_vega_dynamics_2017}. There are some studies to investigate the possibilities to exploit these effects as a resource for quantum information~\cite{huelga_non-markovianity-assisted_2012,chin_quantum_2012}. However, yet there is no trivial conclusion to define these effects as resources. Moreover, they are often reported as detrimental against control techniques~\cite{addis_problem_2016}. 

In this subsection we investigate the dynamics of our  QNN unit connected to a non-Markovian quantum reservoir. In other words, we try to answer the question; What happens when the QNN unit has a memory? To this end, we use a previously reported collision model in order to simulate the dynamics~\cite{cakmak_non-markovianity_2017}. The dynamics is similar to the Markovian case, this time there is an extra coupling between inter-units representing the reservoir as depicted in Fig.~\ref{fig:Fig3} (a). In this scenario, the state of the interacted and the just discarded unit couples with the forthcoming unit in which its $2^{nd}$ nearest neighbour by 

\begin{equation}
H_{uu}=J_{uu}\sum_i \sigma_i^{+}\sigma_{i+2}^{-}+H.c.
\end{equation}
where, $H_{uu}$ is the inter unit coupling Hamiltonian and $J_{uu}$ is the coupling coefficient between units. We couple the QNN unit with one input node, the simplest version as we examined in the Markovian case and again calculate the quantum mutual information during the dynamical evolution. We observe typical revival scheme in the mutual information content (see Fig.~\ref{fig:Fig4} (a)) after a certain reduced level. 

By combining this result with the evolution of fidelity between the state of the output neuron  and the state of the incoming units, one can  infer that by introducing the non-Markovianity to the system, uncertainty increase dramatically. The oscillatory behaviour of fidelity during the evolution retains the state of the output node in a fuzzy state. Moreover, the saturation of the oscillation is far away from the quantum state of the reservoir. We also calculate the evolution of coherence of the output node as $l_1$ norm of coherence explained in Eq.~\ref{eq:norm}. We observe that coherence has a strict oscillatory behaviour in accordance with the fuzzy state of the output system. We can conclude this subsection by interpreting non-Markovianity as an undesirable effect for quantum neural computing purposes according to these findings. 

\section{Conclusion}
 
Summarizing, we examined a QNN unit for closed and open quantum system dynamics. More specifically, we considered an interacting spin model as a quantum version of a neural network unit. First, we examined the QNN unit for closed system dynamics and then  connect to a reservoir by a conventional flip-flop Hamiltonian. We examined the open system dynamics for both Markov and non-Markov regimes for different number of input nodes and different states of information reservoirs. We adopt a collisional model to simulate open system quantum dynamics. 

According to our results for closed system dynamics, QNN unit reveals a highly unstable output state making the dynamical control difficult. We show that when a QNN unit subjected to reservoirs carrying information content, environmental monitoring can be controlled to implement quantum neuro computing tasks and stable outputs of the proposed quantum system can be obtained by weak coupling to the information environments. We examined the system up to three input nodes with various states including the coherent ones. In the presence of Markov reservoirs, steady output states result as a mixture of reservoir states in contact with input nodes.  It's shown that number of input nodes in contact with reservoirs and the coupling strengths can be suitable control parameters for desired output states. Moreover, we find that steady output states can have some amount of coherence enough to reveal pure quantum effects. 

We also find that coupling to non-Markov reservoirs causing memory effects are detrimental for possible quantum neural computing tasks since these effects create uncertain, unstable and fuzzy quantum states. With our findings we can conclude that a possible quantum neuro-computing procedure could be built on weakly coupled quantum systems benefiting reservoir driven dynamics. Our results also show that quantum information environment driven neural quantum systems can also perform mixed state quantum computing tasks.






\end{document}